\documentstyle[twocolumn,aps,pre]{revtex}

\begin{document}
\draft
\title
{Chaotic mixing induced transitions in reaction-diffusion systems}
\author{Zolt\'{a}n Neufeld$^{1}$, Peter H. Haynes$^{1}$ and Tam\'{a}s T\'{e}l$^2$}
\address{$^1$Department of Applied Mathematics and Theoretical
Physics, University of Cambridge\\ Silver Street, Cambridge CB3
9EW, UK\\ $^2$E\"otv\"os University, Department of Theoretical
Physics, H-1518 Budapest, Hungary}

\date{\today}
\maketitle

\begin{abstract}
We study the evolution of a localized perturbation in a chemical 
system with multiple homogeneous steady states,
in the presence of stirring by a fluid flow.
Two distinct regimes are found as
the rate of stirring is varied relative to the rate of the
chemical reaction.
When the stirring is fast 
localized perturbations decay towards a spatially homogeneous
state. When the stirring is slow (or fast reaction) 
localized perturbations propagate by advection
in form of a filament with a roughly constant width and exponentially
increasing length. The width of the filament depends on the 
stirring rate and reaction rate but is independent of the initial perturbation.
We investigate this problem numerically
in both closed and open flow systems and explain the 
results using a one-dimensional "mean-strain" model
for the transverse profile of the filament that captures the 
interplay between the propagation of the reaction-diffusion front
and the stretching due to chaotic advection.
\end{abstract}

\section*{}

{\bf The problem of chemical or biological activity in fluid
flows has recently become an area of active research 
\cite{Menzinger,Epstein,Metcalfe,Merkin,Reigada,Abraham,PNAS,BYoung}.
Apart from theoretical interest this problem has a number
of industrial \cite{Poinsot} and environmental 
\cite{Legras,Haynes} applications.
One of the simplest manifestation of non-linear behavior in
reaction-diffusion systems is the possibility of
travelling front solutions. In this paper we study the effect
of chaotic mixing, by an unsteady laminar flow,
in reaction-diffusion systems supporting front propagation.}\\

\section{Introduction}

Let us consider $N$ interacting chemical or biological
components, with dimensionless concentrations 
$C_i({\bf r},t), i=1,..,N$, advected by an incompressible fluid
flow, ${\bf v}({\bf r},t)$, that is assumed 
to be independent of the concentrations. The spatio-temporal
dynamics of the fields is governed by the set of
reaction-advection-diffusion equations
\begin{equation}
{\partial \over \partial t} C_i+ {\bf v}({\bf r},t) \cdot \nabla C_i =
{\cal F}_i(C_1,..,C_N;k_1,..,k_M) + \kappa \Delta C_i
\label{readdiff}
\end{equation}
where the functions ${\cal F}_i$ describe the interactions between the
different components. These may be chemical reactions or, in the case of biological
populations (e.g. different plankton species), 
they may represent growth, grazing, reproduction, death,
predation etc \cite{SH,Brindley}. The parameters $k_i$ are the reaction rates 
characterizing
the speed of the chemical or biological interactions and $\kappa$ is
the diffusivity, assumed to be the same for all components.
We assume that the flow is laminar (smooth) and time-dependent 
implying chaotic advection \cite{chadvection1,chadvection2,chadvection3}, 
i.e. fluid elements separate exponentially 
at a rate given by the Lyapunov exponent \cite{Ott}, $\lambda$, of the advection dynamics.

The equation (\ref{readdiff}) can be non-dimensionalized by the transformations
\begin{eqnarray}
&x& \to {x \over L}, \;\; t \to {t U \over L}, \;\; v \to {v \over U},
\;\; k_i \to {k_i \over k_1} \nonumber \\
&{\cal F}_i&\left(C_1,..,C_N\right) 
\to {{\cal F}_i(C_1,..,C_N) \over k_1}, 
\end{eqnarray}
where $L$ and $U$ are the characteristic length-scale and velocity of the flow
and one of the reaction rates, $k \equiv k_1$, is used to define a characteristic
chemical time-scale. Thus, the non-dimensional problem can be written as
\begin{equation}
{\partial \over \partial t} C_i+ {\bf v}({\bf r},t) \cdot \nabla C_i =
Da {\cal F}_i(C_1,..,C_N) + Pe^{-1} \Delta C_i
\label{nondim}
\end{equation}
where 
\begin{equation}
Da \equiv {k L \over U} \;\;\;\; {\rm and} \;\;\;\; Pe \equiv {L U \over \kappa}
\label{nondim1}
\end{equation}
are the Damk\"ohler and the P\'eclet number, respectively.
The Damk\"ohler number \cite{Damkohler,Tabeling}, $Da$, 
characterizes the ratio between the advective and the chemical (or biological)
time-scales. 
Large $Da$ corresponds to slow stirring
or equivalently fast chemical reactions and vice versa.
The P\'eclet number, $Pe$, is a measure of the relative 
strength of advective and diffusive transport.
We are going to consider the case of large $Pe$ number,
typical in many applications,
when advective transport dominates except at very small scales.


In applications it may be useful to understand the behavior
of a chemical system for a range of stirring speeds
when other parameters (e.g. reaction rate constants, diffusivity) are kept unchanged.
Although the stirring speed affects both
non-dimensional numbers in (\ref{nondim1}), its variation 
leaves the
family of curves on the $Da - Pe$ plane defined by $Da Pe = {\rm constant}$
invariant.
For this case the appropriate non-dimensional equation could be obtained by
dividing both sides of (\ref{nondim}) by $Da$, implying the use of the 
chemical time-scale, $1/k$, as the time unit.  
Then the two non-dimensional parameters would be $Da$
and $Da Pe = k L^2/\kappa$, where the latter, representing the ratio of the diffusive  
and the chemical time-scales, is independent of the stirring rate.

In this paper,
we study the behavior of the chemical system for different
Damk\"ohler numbers, when the P\'eclet number is kept fixed.
Strictly speaking, this
corresponds to the variation of the chemical reaction rates and cannot be
achieved by changing the stirring rate alone (except if the diffusivity is
also changed in order to keep $Pe$ constant).

For simplicity, in the following we consider the case of two-dimensional flow
only, since it is computationally cheaper and easier to visualize.
This is directly relevant to certain
geophysical problems where density stratification
makes the flow quasi-two-dimensional
(e.g. stratospheric chemistry, plankton in the ocean surface
layer) and is also accessible to laboratory
experiments (using soap-films, stratified fluids etc.). However
we believe that many of the results presented in this paper
can be straightforwardly extended to three dimensional systems.

With the above assumptions the problem defined by Eq.\ref{nondim} 
is still very general until the interaction terms, ${\cal F}_i$ are specified.
Previous work has investigated the spatial structure of
the chemical fields in the case of chemical dynamics
with a single stable local equilibrium concentration, that is a function of
the spatial coordinate \cite{Chertkov1,Chertkov2,PRL1,PRE1}.
Here we consider the case of chemical systems that in the
spatially homogeneous case would have multiple steady state solutions.
There are many examples of multiple steady states in interacting chemical
or biological systems, in models of atmospheric chemistry \cite{Brasseur},
or in the dynamics of plankton populations \cite{SH,Malchow}.
Perhaps the simplest is the auto-catalytic reaction
$A+B \to 2B$.

Thus we assume that the dynamical system that describes
the evolution of the spatially uniform chemical system, 
$C_i({\bf r},t)\equiv C_i(t)$,
\begin{equation}
{d C_i \over d t} = Da {\cal F}_i (C_1,..,C_N)
\end{equation}
has more than one, stable or unstable,
fixed point, or equivalently the system of equations
\begin{equation}
{\cal F}_i (C_1^*,..,C_N^*) = 0
\end{equation}
has multiple roots in the positive quadrant. (Since the concentration
fields must be positive everywhere, only 
the fixed points with $C^*_i \geq 0$ are relevant.)

We study the response of the system, initially in one of the
uniform steady states, to a localized perturbation.
By localized we mean, that the 
the spatial extent of the perturbation, $\delta$, is much smaller
than the characteristic length-scale of the velocity field, $\delta \ll 1$. 
The evolution of the chemical fields is investigated for different values
of the Damk\"ohler number in both closed flow and open flow systems.
The cases of stable and unstable initial uniform states will
be considered separately, each represented by a simple model system.

In the following section we describe the models
used for the reaction dynamics and for the flow,
followed by the presentation of explicit two-dimensional numerical simulations in
section III.
Then, in section IV. we introduce and investigate a one-dimensional 
reduced model and show that this
may be used successfully to interpret the two-dimensional numerical simulations.
The paper ends with a summary and discussion.
 
\section{The models}

For the reaction term, ${\cal F}$, we use two different models 
with multiple equilibria with quadratic and cubic non-linearity.
The first is an auto-catalytic reaction, a generic model
for the propagation of a stable phase into an unstable one. 
The second model is a bi-stable system and we study
the triggering of a transition from one stable state to the other by a 
localized perturbation.
 
1. {\bf Auto-catalytic reaction} -
Let us consider the reaction $A+B \to 2B$,
with the corresponding rate equations for the spatially
uniform system
\begin{equation}
{dC_A \over dt} = - r C_A C_B, \;\; {dC_B \over dt} = r C_A C_B.
\end{equation} 
Observing that $C_A + C_B$ (the total number of molecules $A$ and $B$)
is conserved by the reaction,
we can eliminate $C_A$ and characterize the state of the system with
a new variable $C \equiv C_B/(C_A+C_B)$ ($0<C<1$) representing
the proportion of component $B$, that evolves according to
\begin{equation} 
{dC \over dt} = k {\cal F}(C) \equiv k C (1-C),
\label{autocatalytic}
\end{equation}
where $k=r(C_A+C_B)$.

There are two steady states: $i.)$ $C=0$ (component $A$ only).
This is unstable, since the addition of a small amount
of $B$ leads to the complete consumption of $A$
through the auto-catalytic reaction; $ii.)$
$C=1$ (component $B$ only). This is stable - a small amount
of $A$ is quickly transformed back to $B$.

The temporal evolution of the spatially homogeneous system
can be obtained by integrating (\ref{autocatalytic}) as
\begin{equation}
C(t)= \left [ 1 + \left ( {1-C(0) \over C(0)} \right ) e^{-kt} \right ]^{-1}
\label{transition}
\end{equation}
where $C(0)$ is the initial proportion of component $B$.

In the numerical experiments, the initial state is 
the unstable phase, $C=0$,  
perturbed by a localized pulse of the form
\begin{equation}
C(x,y,t=0)=C_0 e^{-(x^2 + y^2)/2 \delta^2}
\label{perturbation}
\end{equation}
where $C_0 \ll 1$ and $\delta \ll 1$, representing a small
dilute patch of the catalyst, component $B$.

The reaction term (\ref{autocatalytic})
can also be interpreted as the, so called, logistic
growth of a population \cite{Murray} modelling  
the growth of the population limited by the availability of the resources
(e.g. nutrients), implying that the concentration saturates at $C=1$.  
Furthermore, the same reaction term appears in the Kolmogorov-Piskunov-Petrovsky
(KPP) equation \cite{KPP} used in combustion \cite{Poinsot,Ryzhik}, 
which describes the propagation of a flame front
separating fresh (unburned) premixed reactants ($C=0$) and burned gases ($C=1$).
Thus our numerical experiments can also be regarded as representing the
evolution of a plankton bloom stirred by ocean currents or of a flame
embedded in a laminar flow.

2. {\bf Bi-stable system} -
As a second model we use a system with two
stable states defined by the reaction equation 
\begin{equation}
{d C \over dt} = k {\cal F}(C)=k C (\alpha-C) (C-1), \;\;\; 0 < \alpha < 1.
\label{bistable}
\end{equation}
The stable fixed points are $C=0$ and $C=1$, and their basins of
attraction are separated by the unstable fixed point, $C=\alpha$.
Although this system does not represent a
real chemical reaction scheme, it is the simplest 
possible model for multi-stability. We expect that more realistic
chemical and biological systems with multiple stable steady states
would show similar behavior.
We start the system in the stable uniform state, $C=0$, and a localized
perturbation of the form (\ref{perturbation}) is added. 
Now the amplitude of the perturbation is chosen
such to exceed the threshold, $\alpha$, (we use $C_0=1.0$).
(Since the initial state is stable, any perturbation that is below the 
threshold everywhere, $C(x,0)<\alpha$, dies out.)

We note that when $C=\alpha$ is chosen as an initial state
the bi-stable model shows
qualitatively similar behavior to the auto-catalytic model.

Stirring will be modelled by two simple 
time-periodic model velocity fields, representing a closed and an open 
flow system, respectively. We stress, however, that the results described 
in this paper are valid for a wide class of two-dimensional time-dependent laminar 
flows, since the only important assumption is the chaotic
motion of fluid elements.

For the closed flow we choose a sinusoidal shear flow
with the direction of the shear periodically alternating along
the $x$ and $y$ axis \cite{Varosi,Pierrehumbert}. The velocity field is
\begin{eqnarray}
v_x(x,y,t) &=& {A } \Theta
\Bigl({1 \over 2} - t \bmod 1 \Bigr ) \sin{\left ({2 \pi y
}+\phi_i\right)} \nonumber \\
v_y(x,y,t) &=& {A } \Theta
\Bigl(t \bmod 1 - {1 \over 2} \Bigr ) \sin{\left ({2 \pi x
}+\phi_{i+1}\right)},
\label{closedflow}
\end{eqnarray}
defined on a doubly periodic square domain of unit length,
$\Theta$ is the Heaviside step function and
$A$ is a parameter (we use $A=0.7$), that controls the chaotic behavior of the flow.
To avoid transport barriers (due to KAM tori \cite{Ott}, typically present in
periodically driven conservative systems) the 
periodicity is broken by using a random phase, $\phi_i$, different
in each time period.

We also consider open flow systems in which fluid 
continuously flows in and out a finite mixing zone.
Such systems are relevant for
certain chemical reactors and also in some geophysical problems. 
Advection in this type of open flows has been
shown to be governed by a chaotic scattering type escape 
process generating fractal patterns of the advected tracers \cite{open1,openexp}.

As an example of an open flow system we use a velocity field
modelling the flow around two alternately opened point-sinks
in an unbounded two-dimensional domain \cite{Arefsink,karolyi}.
The fluid particles approach the mixing zone from infinity
and leave the domain through one of the sinks.
The velocity field is composed by the superposition 
of a point-vortex and a point-sink, centred on the
active sink. The complex potential corresponding to the vortex-sink is
\begin{equation}
{w}({z})= -(Q+iK)\ln{|z-z_s|}
\end{equation}
where ${z=x+iy}$ is the complex coordinate, $z_s$ is the position
of the sink and $Q$ and $K$ are the
sink-strength and vortex-strength, respectively.
The corresponding velocity field in polar coordinates with the
origin fixed to the active sink is
\begin{equation}
v_r = - {Q \over r}, \;\;\;, v_{\phi} = {K \over r}, \;\; 
r=\sqrt{(x - x_s)^2 + (y - y_s)^2}.
\label{openflow}
\end{equation}
The half distance separating the two sinks
is assumed to be unity ($x_s = 0, y_s= \pm 1$). The sinks are alternately opened
for equal times and the period of the flow is the time unit.
The inflow concentration at the boundaries of a square domain
is kept constant at 
the value corresponding to the initial background concentration, 
$C(\pm 3.0,y)=C(x,\pm 3.0)=0.0$.

For both closed and open flows we integrate the advection-reaction-diffusion 
problem for the auto-catalytic and the bi-stable model 
on a $1000 \times 1000$ lattice using a semi-Lagrangian scheme. The value of 
the P\'eclet number is $Pe = 10,000$, and the Damk\"ohler number is varied in a
range from zero to few hundred.

\section{Numerical results}

\subsection{Closed flow}

We find for both chemical models two distinct regimes
separated by a critical Damk\"ohler number, $Da_c$.
The critical values are $Da_c \approx 2.0$ for the auto-catalytic reaction
and $Da_c \approx 20.0$ for the bi-stable model.

In the slow reaction/fast stirring regime
($Da<Da_c$), in both models the initial perturbation quickly decays towards
a homogeneous state (Fig.1).  
In the case of the auto-catalytic reaction the homogenisation of the 
perturbation is followed by a spatially uniform transition to the
stable equilibrium, $C=1$, as in a reactor with initially premixed components. 
This is because the original
uniform state is unstable, and cannot be restored by the homogenisation
of the perturbation, since the homogenised state still deviates slightly
from the unstable equilibrium. In the bi-stable system the perturbation
is dispersed and the
system simply returns to the unperturbed initial state, $C=0$.
In this regime, the chemical reaction is too slow to sustain
the localized perturbation that is diluted by the
strong stirring. Note that for both chemical models the final state would be the same  
for a spatially uniform perturbation with the same
spatial mean. Thus, a coarse grained model could, at
least qualitatively, reproduce the evolution of the system.

When $Da>Da_c$ the localized perturbation may persist and propagate
in the form of a filament with a roughly constant width and
rapidly increasing length (Fig.2). This continues until the 
filaments cover the whole domain and finally the system 
becomes uniform again, $C=1$.
This occurs in all cases in the auto-catalytic case, but only if
$\alpha < 0.5$ in the bistable case. If $\alpha > 0.5$ a localised 
perturbation in the bistable case decays.
The width of the propagating filaments increases with $Da$.
We note, that the average profile of the filament (width, amplitude) 
is apparently independent of the details of the initial perturbation,
indicating that
it is determined by the interplay between the chemical and transport
dynamics in the system.
In this regime the transition from the initially uniform state
to the final one is strongly non-uniform in space. 
Since the spatial variation is essential, a coarse grained
model that was unable to resolve the filaments, 
would produce very different outcomes.

The transition from the non-homogeneous to homogeneous reaction in the
case of the auto-catalytic model for decreasing $Da$ is clearly shown by the 
snapshots of the spatial distribution taken at the midpoint of the transition
defined by the spatial mean concentration $\bar C =0.5$ (Fig.3.).
To characterize the change in the non-uniformity of the reactions as
$Da$ is varied we plotted the relationship between the first and
the second moments ($M_1 \equiv \langle C \rangle$ and $M_2 \equiv \langle C^2 \rangle$)
of the chemical distribution (Fig.4).
There are two extreme situations. In case of a spatially uniform system
the averaging can be ignored and thus $M_2 = M_1^2$.
For a strongly non-uniform distribution with only two
possible values $1$ and $0$ (representing a two-phase system with a
very narrow transition zone) the square for the second moment is
irrelevant and $M_2 = M_1$.
Fig.4. clearly shows the transition from the linear to the quadratic
relationship as $Da$ is decreased.

The time-dependence of the mean 
concentration, $\bar C $, (equivalent here to the
spatial average of the deviation from the initial state)
is shown in Fig.5 for different $Da$ numbers.
For large $Da$ numbers, after an initial transient time, 
an exponential growth can be observed with the growth rate
independent of the Damk\"ohler number for both chemical models.
This shows that the growth of the mean concentration
is controlled by the stirring, that increases 
the length of the propagating filament.
Note, that for the bi-stable system a localized perturbation
with its spatial mean concentration well below the
threshold, $\alpha$, can flip 
the system to the other steady state. This is a strong example
where spatial smoothing does not work.

For $Da<Da_c$ the homogeneous dynamics is relevant.
In the bi-stable system the mean concentration simply decays 
to zero exponentially as in the homogeneous system (for $\bar C \ll 1$)
\begin{equation}
{d \bar C \over dt} = Da {d {\cal F} \over d \bar C} \bar C = - \alpha Da  \bar C,
\;\; \bar C \sim e^{-\alpha Da t}
\end{equation}
In the auto-catalytic model,
the growth of the mean concentration depends on the
$Da$ number. When the time-dependence is plotted against $Da t$ (Fig.6), 
(i.e. the time unit is the chemical time) the
curves corresponding to $Da<Da_c$ collapse 
showing that in this regime the 
transition is independent of the stirring rate, as expected for
a spatially uniform system, and the numerical results agree
well with the solution obtained for the homogeneous system (\ref{transition}). 
In this regime the two reactants
(A and B) are brought close to each other by the flow at a higher rate than
they can react, therefore further increase of the mixing rate cannot
enhance the production.
The growth of the mean concentration is limited by mixing
(transport) for supercritical Damk\"ohler numbers and limited by the
chemical reaction (local dynamics) below the critical value.


\subsection{Open flow}

In the open flow case we find again a transition at a 
critical value of the Damk\"ohler number.

When the stirring is strong ($Da<Da_c$) any localized perturbation 
dies out and both models return to the original
state.  There is no homogeneous transition
for the auto-catalytic system as this would be incompatible
with the inflow boundary conditions. In other words, 
the perturbation is expelled completely from the system 
(through the sinks)
and thus even the unstable basic state ($C=0$) can be restored.

When $Da>Da_c$, similarly to the closed flow case, the perturbation
produces a propagating filament. However, in this case due to the continuous
outflow and inflow the filament cannot fill the domain uniformly.
After a short transient 
a non-uniform stationary state sets in, with the mixing zone partly
covered by a complex filamental structure (Fig.7). In 
our case the stationary pattern varies periodically
with the period of the flow. 
For the auto-catalytic model, a small amplitude
perturbation is sufficient to initiate the propagating filament, 
while in the bi-stable model only perturbations 
larger than the threshold, $\alpha$, are able to trigger
the transition to the non-uniform stationary state.
(Furthermore an additional condition for the existence of
the non-uniform stationary state is again that $\alpha<0.5$.)

One can observe that the non-uniform (periodic) stationary pattern
follows the fractal unstable manifold of the non-escaping set formed by fluid
particles that are trapped forever in the mixing zone \cite{karolyi}.
The unstable manifold can be easily visualized by simply following the evolution 
of an ensemble of fluid elements (e.g. representing a droplet of dye) advected 
by the flow (Fig.8).
For the auto-catalytic reaction this has been
already observed in a kinematic model where B particles are treated
as individual tracers \cite{openc}.
As in the closed flow case, the width of the filaments
increases with the $Da$ number (Fig.9).
The dependence on the $Da$ number
of the total concentration (i.e. the
area covered by the filaments) in the stationary state 
is shown in Fig.10. We find a continuous transition for the auto-catalytic 
reaction ($Da_c = 2.3$) and a discontinuous one for the bi-stable model ($Da_c=24.2$). 

\section{The Lagrangian filament slice model}

Here we introduce a reduced one-dimensional model in order to 
explain the numerical results described in the previous section.
In the presence of chaotic transport fluid elements are stretched 
into elongated filaments. This is well known
from numerical simulations and has been observed in 
laboratory experiments using dye droplets \cite{chadvection3,openexp,closedexp}.
In a two-dimensional system, one can 
assign to any point of the flow
a convergent and a divergent direction associated with the 
eigenvectors corresponding to the negative
and positive Lyapunov exponents ($-\lambda$ and $\lambda$)
of the chaotic advection. These directions are tangent to the
stable and unstable foliations of the advection dynamics, respectively.
Any advected material line (e.g. iso-contours of a conserved tracer)
tends to align along the unstable
foliation in forward time, or along the stable foliation in backward time.
Thus, the stirring process smooths out the concentration of the advected tracer
along the stretching direction, whilst enhancing the concentration gradients
in the convergent direction.

Let us now separate the original reaction-advection-diffusion
problem along the (Lagrangian) stretching and convergent directions.
In the stretching direction the perturbation is spread by
advective transport, that is the dominant process being 
much faster than diffusion. In the convergent direction the
formation of small scale structure indicates that all 
three processes of reaction, advection and diffusion
are important and need to be considered together.
The resulting equation determines the mean transverse profile of the filament
that propagates along the divergent direction following the
unstable foliation.

Thus one expects that the locus of the centre of the filament can be obtained 
simply by advecting a material contour starting in the region 
of the initial perturbation. 
This is confirmed by the numerical simulations as shown in Fig.11 for the
closed flow model, and is consistent with Fig.8 for the open flow system.
An important difference 
between the two is that while in the closed flow the contour gradually fills
the whole domain, in the open flow it draws out the unstable manifold
of the set formed by all non-escaping orbits. 
The length of the contour increases exponentially (Fig.12), $L(t) \sim \exp{(\theta t)}$
with $\theta \approx 2.05$. 
We note, that the contour lengthening rate, $\theta$ is always larger than the
Lyapunov exponent ($\lambda$), which represents the average growth rate of 
infinitesimal
line elements. This is because the instantaneous stretching rate fluctuates
and the increase of the total length is dominated by the growth of
a line elements that experience a faster than average stretching.
In dynamical systems language the contour lengthening
rate $\theta$ is given by the topological entropy \cite{Ott,Newhouse,Alvarez} 
of the advection dynamics.

In the convergent direction we have the following one-dimensional 
equation for the average profile of the filament  
\begin{equation}
{\partial \over \partial t} {C}- \lambda x {\partial \over \partial
x} {C} = k {\cal F}(C)+ \kappa {\partial^2 \over
{\partial x}^2} {C},
\label{1dmodel} 
\end{equation}
representing the evolution of a transverse slice of the filament
in a Lagrangian reference frame (i.e. following the motion of a
fluid element).
The second term on the left hand side is a stretching term that
takes into account the mean convergent flow
and ${\cal F}$ is the original reaction term.
The stretching term in equation (\ref{1dmodel}) can be interpreted as
advection by a pure strain flow along its convergent direction $v_x=-\lambda x$.
In the two-dimensional problem the strength of stretching
fluctuates in space and time. To capture the average behavior 
this can be represented by a constant stretching rate, $\lambda$, equal to the 
Lyapunov exponent of the chaotic advection.
Therefore the equation above can be regarded as a Lagrangian mean 
field description. Equation (\ref{1dmodel}) has also been studied recently
by McLeod et al \cite{McLeod} investigating the filament width of oceanic plankton
distributions.

The equation (\ref{1dmodel}) is defined on $-\infty < x < \infty$
with the boundary conditions 
\begin{equation}
C(x \to \pm \infty) = 0, \;\;\; {d C \over dx}(x \to \pm \infty) = 0.
\label{1dboundary}
\end{equation}
representing the assumption that most of the system is in the
background state, $C=0$, so that different parts of the filament
are well separated from each other and they do not interact.
The single filament approximation is not valid for the late
stage of the evolution when the filaments can overlap.

The homogeneous steady state, $C(x) \equiv 0$, is a trivial solution
of (\ref{1dmodel}-\ref{1dboundary}).
Let us now consider the evolution of a localized pulse-like disturbance.
(We use perturbations centred at the origin, but it turns out that the 
asymptotic behavior is independent of the initial position of the perturbation.)  

For a non-reactive tracer ($k=0$) the equation (\ref{1dmodel}) has asymptotic
solutions (for large $t$) of the form of a Gaussian pulse whose amplitude decays 
exponentially in time
\begin{equation}
C(x,t) \sim  \exp{(- \lambda t )}
\exp{ \left ( -{x^2 \lambda \over 2 \kappa } \right )}.
\end{equation}
The width of the Gaussian, $l_{dif} \equiv  \sqrt{\kappa / \lambda}$,
is determined by the balance
between the strain and diffusion.

Let us now consider the case of a reactive tracer ($k \neq 0$)
without stretching ($\lambda=0$).
It is well known that reaction-diffusion systems with
multiple equilibria have travelling front solutions connecting
different steady states \cite{Murray,AScott,Keener}. The fronts move with
a constant speed $v_0$ with no change of shape, $C(x,t)=C(x-v_0 t)$.

The reaction-diffusion problem corresponding to the auto-catalytic
model is known as the Fisher equation \cite{Fisher} 
(or Kolmogorov-Petrovsky-Piskunov equation \cite{KPP} in the combustion literature) 
and describes the propagation
of a stable phase ($C=1$, component B) into an unstable one ($C=0$, component A). 
Localized perturbations
evolve into a pair of fronts moving away from the centre
with the asymptotic speed, $v_0=2\sqrt{k \kappa}$.


For the bi-stable model (\ref{bistable}) the velocity of the front 
joining the two stable states, $C(x \to -\infty)=1$ and $C(x \to \infty)=0$, is \cite{Murray,AScott,Keener}
\begin{equation}
v_0 = \sqrt{k \kappa} \int_{0}^{1} {\cal F}(C) dC =  
\sqrt{k \kappa \over 2}(1-2\alpha).
\end{equation}
The sign of the above expression can be either positive
or negative showing that the direction of the propagation depends 
on the parameter $\alpha$. The single travelling front solution can
be found analytically \cite{Murray} as
\begin{equation}
C(x-v_0 t) = \left [ 1 + \exp{ \left ({x - v_0 t \over \sqrt{2}} \right )}
\right ]^{-1}.
\end{equation}
When the initial basic state is $C=0$ a localized
perturbation can initiate a pair of propagating fronts moving
away from each other only if
$\alpha<0.5$, otherwise the direction of the front propagation is such
that the fronts approach each other and the perturbation dies out.
This explains the decay of the perturbations for $\alpha>0.5$
in the two-dimensional simulations independently of the stirring
rate.

Thus, in the absence of stretching both type of systems have travelling 
front solutions with the front speed proportional to $\sqrt{k \kappa}$.
A localized perturbation generates a pair of fronts moving
in opposite directions away from the centre. (For bi-stable systems
this happens only if $\alpha <0.5$).
For the auto-catalytic model the amplitude of the perturbation
can be arbitrarily small, while it must exceed the threshold $\alpha$ in the
bi-stable case. 

In the presence of stretching, $\lambda > 0$ one expects that 
the convergent flow will eventually stop the propagation of the fronts
at a point $x=w$ where the speed of the propagation is balanced by
the advection
\begin{equation}
w \lambda \approx \sqrt{k \kappa}
\label{filwidth1}
\end{equation}
This gives an estimate for the width of the resulting 
filament as
\begin{equation}
w \sim {\sqrt{k \kappa} \over \lambda} = l_{dif} \sqrt{ \tilde {Da}}, 
\;\; \tilde {Da} \equiv {k \over \lambda}
\label{filwidth2}
\end{equation}
where we have introduced the Lagrangian Damk\"ohler number, $\tilde {Da}$.
This is defined on the basis of the Lyapunov time, $1/\lambda$,
of the flow, representing the Lagrangian characteristic time-scale 
of the two-dimensional advection problem. This turns out to be 
more appropriate for the filamentation problem than the definition (\ref{nondim1})
based on Eulerian characteristics like the average flow velocity $U$.

The propagation velocity (\ref{filwidth1}) is for an isolated
front only. Therefore we expect that the scaling for the filament width 
(\ref{filwidth2}) is
valid when the distance between the two fronts, representing
the edges of the steady filament, is sufficiently large compared to the 
diffusive scale, $w \gg l_{dif}$, that is $\tilde {Da} \gg 1$.

Equation (\ref{1dmodel}) can be non-dimensionalized by
using the Lyapunov time, $\lambda^{-1}$, as the time-scale unit
and the diffusive scale, $l_{dif}$, as the unit length
\begin{equation}
{\partial \over \partial t} {C}- x {\partial \over \partial
x} {C} = \tilde {Da} {\cal F}(C)+ {\partial^2 \over
{\partial x}^2} {C}.
\label{nondim1d}
\end{equation}
Since the one-dimensional problem is defined on an unbounded domain,
the P\'eclet number does not appear in (\ref{nondim1d}).
For the two-dimensional problem the characteristic scale of the
velocity field, $L$, is finite and this can be used to define
a Lagrangian P\'eclet number based on the Lyapunov exponent of the flow as
\begin{equation}
\tilde {Pe} \equiv { L^2 \lambda \over \kappa} = \left ( {L \over l_{dif}} \right )^2.
\end{equation}
Using this the expression for the width of the filament can be rewritten as
\begin{equation}
{w \over L} \sim {\sqrt{k \kappa} \over \lambda L} = \sqrt{ \tilde {Da} \over 
\tilde {Pe}}.
\label{2dfilwidth}
\end{equation}
The straining flow approximation can only be valid for scales smaller
than $L$ ($w \ll L$), thus (\ref{2dfilwidth}) is expected to be correct
in the range $1 \ll \tilde {Da} \ll \tilde {Pe}$.

In the open flow model in the stationary state
the filaments cover a fractal set. For the parameter values used
in our simulations the dimension of this was found to be $D \approx 1.69$.
The area of the large concentration region can be obtained by using
the dimensionless filament width $w/L$ as the resolution when observing this region.
Since the number of boxes of size $w/L$ needed to cover the region is proportional to
$(w/L)^{-D}$ \cite{Ott}, the area $A$ of this regions is
\begin{equation}
{A \over L^2} \sim { \left ({w \over L} \right )}^{2-D} = 
{ \left ( {\tilde {Da} \over \tilde {Pe}} \right ) }^{(2-D)/ 2}.
\end{equation}
This shows, that due to the overlaps the total area grows more slowly with the  
Damk\"ohler number than the area of a single isolated filament.
An analogous scaling has been obtained for auto-catalytic reaction
on an open baker map in \cite{baker}.

Numerical simulations of the reduced problem
(\ref{nondim1d}) show that for both chemical models there is
a critical value of the Damk\"ohler number. When $\tilde {Da}> \tilde {Da}_c$
there exists a non-uniform steady solution to (\ref{nondim1d}) centred 
on the origin (Fig.13).
Otherwise all perturbations
decay and the only steady solution is the trivial one, $C(x) \equiv 0$. 
The width of the non-uniform steady solution increases with $\tilde {Da}$,
that appears to be consistent with (\ref{filwidth2}) for large Damk\"ohler numbers.
Numerical continuation of the non-uniform steady solution for 
decreasing $\tilde {Da}$ confirms that this solution disappears at 
a critical value.
The transition is found to be continuous for the auto-catalytic 
model and discontinuous for the bi-stable model (see Fig.13).
Let us now analyse the transition in the reduced problem (\ref{nondim1d})
in more details for the two chemical models, separately.

\subsection{Auto-catalytic model}

For the auto-catalytic model the non-uniform solution approaches and 
coalesces with the uniform solution when $\tilde {Da}_c$ is approached 
from above.
When $\tilde {Da}< \tilde {Da}_c$ localized perturbations decay, showing that
the uniform state, $C(x) =0$, in spite of being an unstable fixed point
of the homogeneous system is stable against localized disturbances.
Thus stirring stabilizes the unstable equilibrium of the homogeneous system
by dispersing and diluting the perturbation.
This is consistent with the behavior observed in the two-dimensional
simulations for sub-critical Damk\"ohler numbers, showing homogenisation and
decay of the perturbation followed by growth only after the reactants were distributed
uniformly in space.
For supercritical Damk\"ohler numbers the uniform solution is unstable
against localized perturbations.
Arbitrarily weak perturbations are sufficient to reach the non-uniform steady state.

Just above the critical point, $\tilde {Da}-\tilde {Da}_c \ll 1$,
the amplitude of the non-uniform solution is small and 
the chemical dynamics can be linearized about the background state
\begin{equation}
{\partial C\over \partial t} - x {\partial C \over \partial
x}  = \tilde {Da} C + {\partial^2 C \over \partial x^2}.
\label{linearFisher}
\end{equation}
This problem has been studied by Martin \cite{Martin} in the context of plankton
populations.
Equation (\ref{linearFisher}) has asymptotic solutions of the form
\begin{equation}
C(x,t) \sim  e^{-x^2/2 } e^{(\tilde {Da}-1)t},
\end{equation}
that decay in time when $\tilde {Da}<1$ and grows 
exponentially, moving out of the domain of validity of the linear 
approximation, otherwise. 
(The non-linearity would eventually stop the growth.)
Thus we obtain the critical value for the auto-catalytic model, $\tilde {Da}_c=1$.

Close to the transition, $\tilde {Da} = 1+\epsilon, 0 < \epsilon \ll 1$, 
one can look for a steady non-uniform solution of the form
\begin{equation}
C(x; \epsilon)=\epsilon C_1(x) + \epsilon^2 C_2(x) + ...
\end{equation}
Substituting this into (\ref{1dmodel}) for
the term first order in $\epsilon$ we obtain
\begin{equation}
{d^2 C_1 \over dx^2} + x {d C_1 \over dx} + C_1 = 0
\end{equation}
that has the solution $C_1 = A e^{-x^2/2}$,
where $A$ is a constant that can be
determined from the equation for the terms second
order in $\epsilon$
\begin{equation}
{d^2 C_2 \over dx^2} + x {d C_2 \over dx} + C_2 = C_1^2 - C_1.
\end{equation}
The left hand side of the equation can be written as 
\begin{equation}
{d \over dx}\left ( x C_2 + {d C_2 \over dx}\right) = C_1^2 - C_1. 
\end{equation}
Integrating both sides over the whole domain
the left hand side vanishes, according to (\ref{1dboundary}),
and an equation for the constant $A$ is obtained
\begin{equation}
\int_{-\infty}^{\infty} (C_1^2 - C_1) dx = A^2 \sqrt{\pi} - A \sqrt{\pi \over 2} = 0
\end{equation}
that gives $A= 1/\sqrt{2}$. Thus, the steady solution
close to the transition point can be approximated as
\begin{equation}
C(x;\epsilon)  = {\epsilon \over \sqrt{2}} e^{-x^2/2} + {\cal O}(\epsilon^2)
\end{equation}

\subsection{Bi-stable model}

For the bistable case the transition at $\tilde {Da}_c$ is discontinuous.
The non-uniform solution disappears with a finite amplitude
far from the uniform state. The uniform solution is stable for
any $\tilde {Da}$ thus small perturbations decay independently 
of the Damk\"ohler number.
In the supercritical regime the uniform and non-uniform stable 
solutions coexist suggesting 
the presence of a threshold for exciting the non-uniform perturbation.

To investigate the transition further let us look for 
steady solutions of eq.(\ref{nondim1d}) 
\begin{equation}
{d^2 C \over dx^2} = - \tilde {Da} {\cal F}(C) - x {d C \over dx}
\label{particle}
\end{equation}
that are consistent with the boundary conditions (\ref{1dboundary}).
Since the non-uniform steady state is symmetric about the origin
it is sufficient to consider the domain $0 < x < \infty $, with 
the constraints 
\begin{equation}
C(x \to \infty) \to 0, {d C \over dx}(0)=0.
\label{particleorbit}
\end{equation}
Note that, if $x$ is interpreted as time and $C$ as a
spatial coordinate, the problem (\ref{particle}) is equivalent to 
the motion of a particle in an asymmetric two-humped potential (Fig.14) 
\begin{eqnarray}
{d V \over dC} &\equiv& \tilde {Da} C (\alpha -C)(C-1),\nonumber \\
V(C) &=& -\tilde {Da} C^2 \left ( {1 \over 4}C^2 + {(1+\alpha) \over 3}C
- {\alpha \over 2} \right ),  
\end{eqnarray}
under the effect of linear friction,
with friction coefficient increasing linearly in time.
The two maxima of the potential
are at the stable fixed points $C=0$ and $C=1$ and 
the minima is at $C=\alpha$.
The potential difference between the two maxima is proportional
to $\tilde {Da}$. The particle trajectory
satisfying (\ref{particleorbit}) starts from the left slope of the higher
potential 'hill' ($\alpha<C(x=0)<1$) with zero velocity and 
ends exactly on the top of the lower hill.
Thus the problem reduces to finding the appropriate values of
the initial coordinate, $C_0 \equiv C(x=0)$.
For initial conditions $C_0 \in (\alpha,1), C'(x=0)=0$ 
the trajectory of the particle may either end in the potential well 
$C=\alpha $ or may cross the smaller hill and escape to $-\infty$.
The trajectories corresponding to the non-uniform
steady solutions are at the boundary between these two
types of asymptotic behavior.

We calculated numerically trajectories for a set of initial
conditions in the range $C_0 \in (\alpha,1)$ for a set of different values of the 
Damk\"ohler number, $\tilde {Da} \in (0,40)$. 
The asymptotic behavior of the trajectories 
is indicated on the $C_0 - \tilde {Da}$ plane (Fig.15):
blank, $C(x \to \infty) \to \alpha$; black, $C(x \to \infty) \to -\infty$.
The required steady solutions are on the boundary
of the two regions. The numerical results show that for
small $\tilde {Da}$ there is no such boundary and a solution of the type
(\ref{particleorbit}) does not exist. In the particle analogy the
interpretation of this is that the difference in the height of the
two hills is not sufficiently large to compensate for
the energy dissipated by friction, thus the particles cannot escape.
When $\tilde {Da}$ is increased above the critical value $\tilde {Da}_c$ 
there are two solutions.
If the hill at $C=1$ is high enough there are initial conditions
for which the particles
have sufficient energy to cross the potential barrier.
Clearly, particles with initial conditions below the height
of the lower potential hill are still
trapped in the potential well. Also, particles 
started from a point very close to the top of the higher hill
are unable to escape since they may
spend very long time in the neighbourhood of the stationary
point and go through the potential well at a late time when 
the friction is strong.
Thus, for $\tilde {Da}< \tilde {Da}_c$ the initial conditions $C_0^{escape}$
for which particles escape to infinity are in an the interval of the form
\begin{equation}
\alpha < C_{0,1}(\tilde {Da}) \leq C_0^{escape} \leq C_{0,2}(\tilde {Da}) < 1
\end{equation}
where $C_{0,1}(\tilde {Da})$ are $C_{0,2}(\tilde {Da})$ are initial conditions
corresponding to steady non-uniform solutions of (\ref{1dmodel}).
As the Damk\"ohler number is decreased the two solutions approach each other 
and disappear at $\tilde {Da}_c = 11.0$.

The trajectory starting from $C(x=0)=C_{0,2}(\tilde {Da})$ clearly corresponds 
to the steady non-uniform solution found in the numerical simulations 
of the time-dependent problem. The solution corresponding
to the lower branch, $C(x=0)=C_{0,1}(\tilde {Da})$, however, 
is not found as an attractor of the time-dependent
problem. This suggests that this is an unstable solution
playing the role of a threshold separating the basins of
attraction of the uniform and non-uniform stable solutions.
(It can be shown that all initial conditions that are above
(below) this separating solution everywhere, converge to the non-uniform
(uniform) steady state. This, of course, does not say anything
about initial conditions partly below and partly above the separating
solution.)

\section{Summary and discussion}

The one-dimensional Lagrangian filament model (\ref{1dmodel}) clearly explains 
the qualitative features
of the two-dimensional numerical results. It shows how a
steady filament profile can arise as a result 
of the interaction between the propagation of
a reaction-diffusion front and stretching due to 
chaotic advection. The disappearance of the filament type
solution for sub-critical Damk\"ohler numbers explains the transition
observed in the two-dimensional simulations.
The advective propagation of the filament along the unstable foliation of the chaotic
advection explains the exponential growth of the mean concentration.
In principle, these results could be used as a numerical technique for obtaining 
the approximate spatial distribution of the chemical components
by combining a  
two-dimensional contour-advection calculation with the information about
the filament width obtained from the steady solution of the one-dimensional
model. 

In order to compare the critical Damk\"ohler numbers predicted 
by the one-dimensional mean strain model with the ones obtained from
the direct numerical simulations we calculate the Lagrangian Damk\"ohler
numbers corresponding to the two-dimensional problem. The Lyapunov
exponents of the advection in the two model flows was found to be:
$\lambda_{closed} \approx 1.66$ and $\lambda_{open} \approx 2.19$.
The critical values of the Lagrangian Damk\"ohler number based on these
Lyapunov exponents are presented in Table 1. and show a very good agreement.
  
In our analysis we neglected the fluctuations of the 
stretching rate. In reality there is a distribution of stretching rates. 
The effect of this is visible in the numerical simulations showing
that the width of the filament slightly fluctuates in space and time.
Also the direction of the stretching fluctuates and foldings of the filament 
may lead to large curvatures whose effect is not captured by our one-dimensional 
description. Another effect is the non-uniform density of
the unstable foliation pointed out by Alvarez et al \cite{Alvarez}.
Thus the advected filament can overlap with itself well before
it fills the whole domain.
Some regions of the flow are filled while others are still empty. 

Here we investigated only reactive systems described by
the distribution of a single species. 
We expect, however, that the basic phenomena
described in this paper remains valid for
multi-component reactive systems that may
have a number of different chemical time-scales. 
One example of this type is the case of excitable systems \cite{Murray,Keener,Meron}
under stirring by a chaotic flow discussed
in \cite{PRL2}. Excitable systems have two different time-scales - 
corresponding to fast
and slow components. Although these systems only have a single
(stable) steady state, the 'rest state', they also have a 'meta-stable'
excited state that persists for a finite time only.
Excitable systems under stirring exhibit similar behavior to the one presented
here, including advective propagation in form of a steady excited filament
and the existence of a critical Damk\"ohler number. 
The one-dimensional excitable reaction-diffusion systems have travelling pulse
solutions, that in the presence of stretching leads to the
appearance of a steady excited filament solution. This can be simple unimodal,
as in our case, but it can also have a bi-modal structure with the central 
part of the filament returned to the rest state.
Thus the existence of an extra chemical time-scale in this system allows for 
somewhat more complex structures and a further
transition in the large $Da$ number range, being a transition
from coherent to non-coherent excitation of the system.

A nice example of an advectively propagating perturbation, of the kind
described in this paper, has been observed recently in a so-called 'ocean
fertilization' experiment \cite{Abraham2}, where the injection of a trace 
element affecting the plankton ecosystem triggered a phytoplankton bloom
in the form of an elongated filament, observed on satellite
images. The response of plankton ecosystem models to such perturbation in
the presence of stirring has been studied in \cite{GRL}.
We suggest that similar phenomena could also be investigated in
laboratory experiments.

\acknowledgements

This work was supported by the UK Natural Environment Research Council
(NERC UTLS Ozone grant NER/T/S/1999/00103) and by the Hungarian
Research Foundation (OTKA T032423) and the MTA-OTKA-NSF Fund (no 526).


\begin{table}[!hbp]
\begin{tabular}{|c|c|c|c|}
& closed & open & 1D model\\
\hline
auto-catalytic & $\approx 1.2$ & 1.05 & 1.0 \\
\hline
bi-stable & $\approx 12.0$ & 11.06 & 11.0 \\
\end{tabular}
\caption{
The critical values of the Lagrangian Damk\"ohler number
for the two-dimensional simulations and for the 
one-dimensional single filament model}
\end{table}

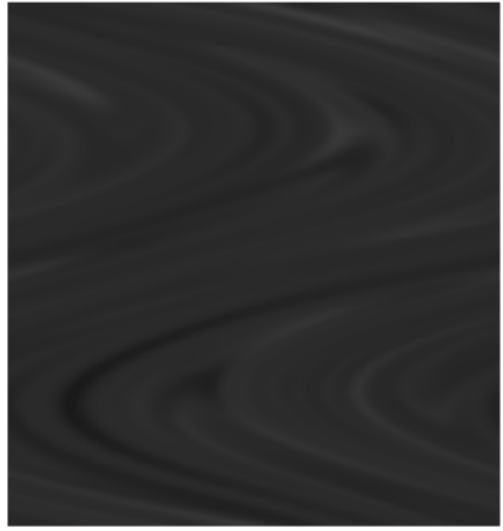
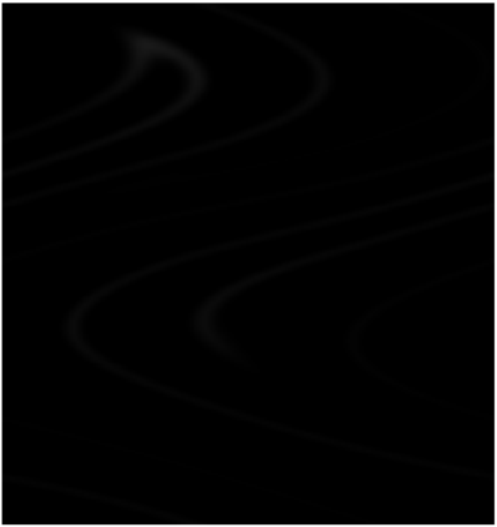
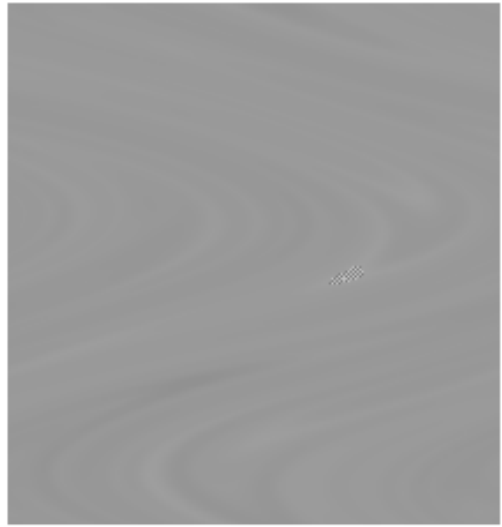
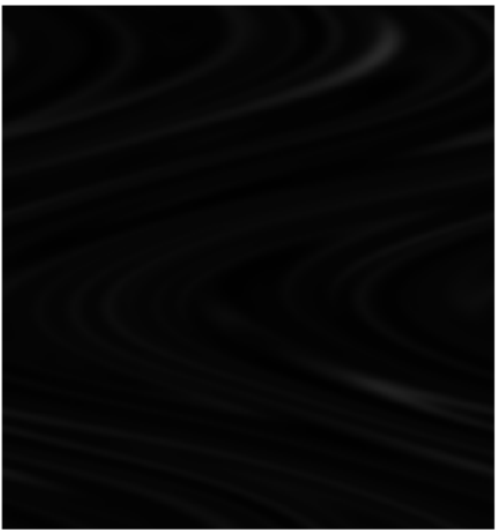
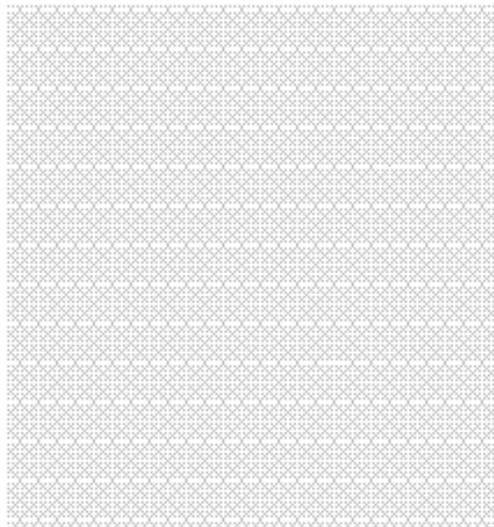
\begin{figure}
\caption{Snapshots of the spatial distribution for the
autocatalytic reaction for $Da=1.0, (< Da_c)$ at $t=0.0, 2.0, 4.0, 6.0, 8.0$.
The amplitude of the initial perturbation is chosen to be
$C_0=0.5$ in order to make the initial decay visible.}
\end{figure}

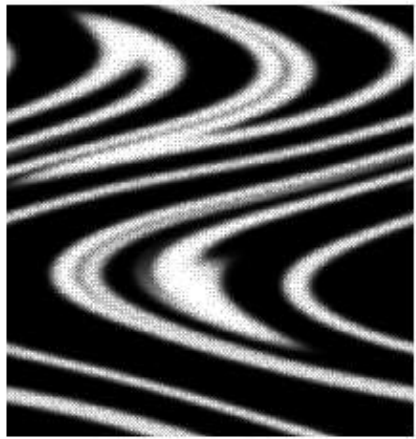
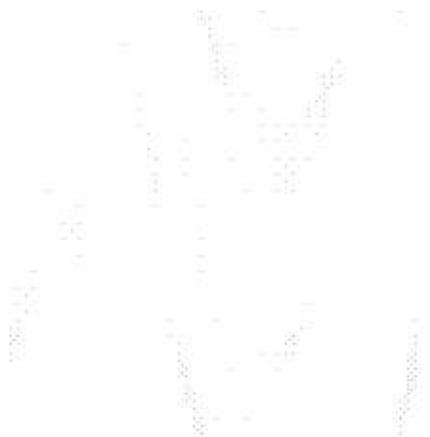
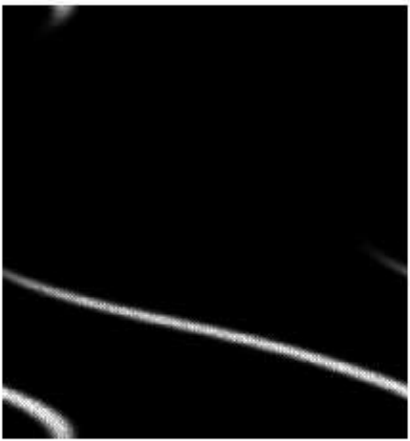
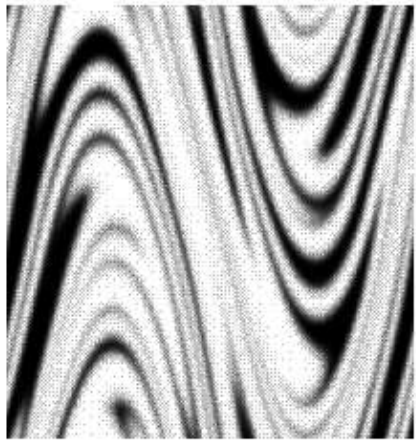
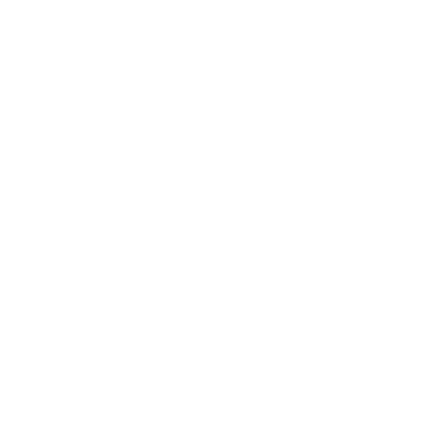
\begin{figure}
\caption{Snapshots of the spatial distribution for the 
autocatalytic model for a supercritical Damk\"ohler number,
$Da=7.0$ at $t=0.0, 0.5, 1.0, 1.5, 2.0, 2.5, 3.0, 3.5$.
The amplitude of the initial perturbation is $C_0=0.5$.
The bistable model shows qualitatively similar behavior.}
\end{figure}

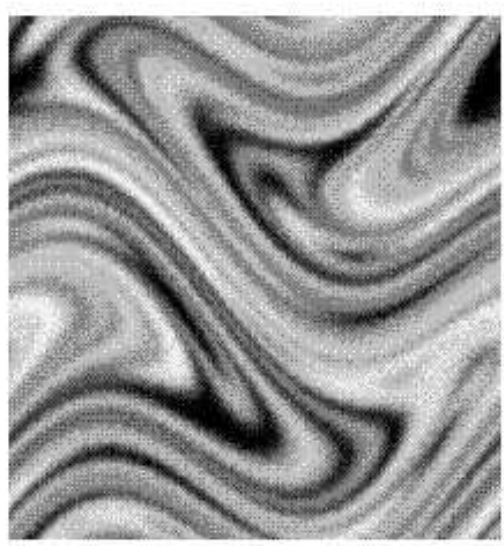
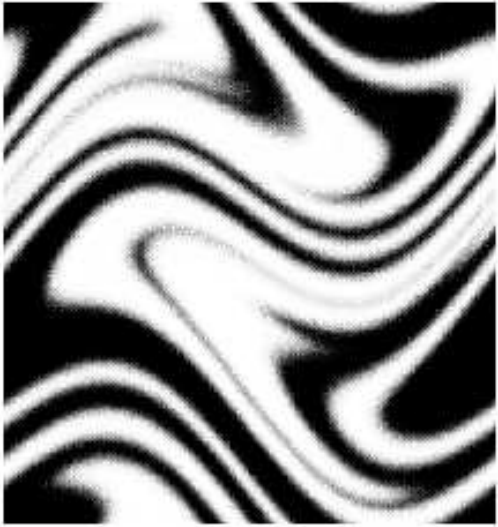
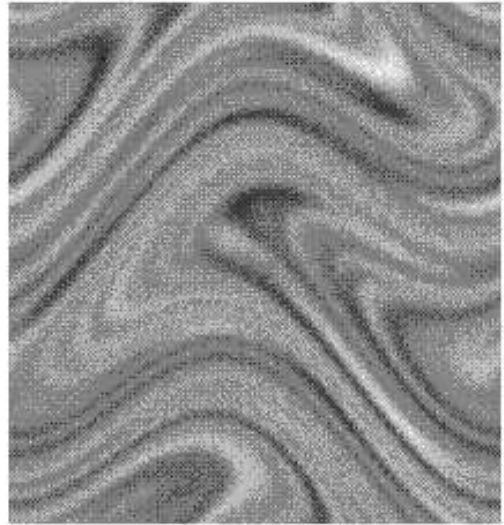
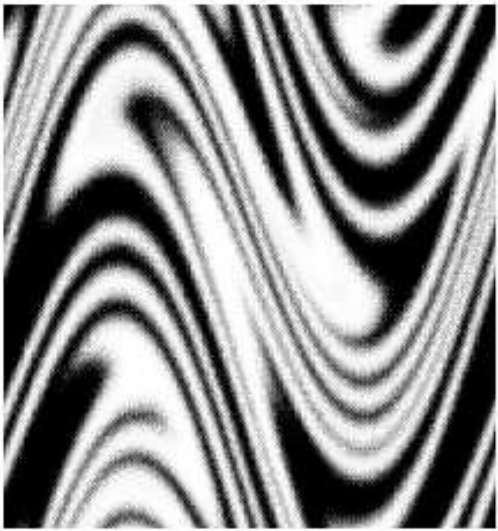
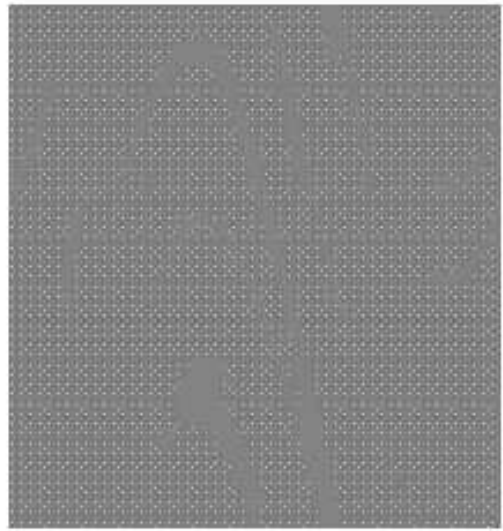
\begin{figure}
\caption{Snapshots of the spatial distribution for the autocatalytic
model at the midpoint of the transition (defined by $<C>(t)=0.5$) for
$Da=35.0, 12.0, 8.40, 4.1, 2.9, 1.0$.}
\end{figure}

\begin{figure}
\caption{The relationship between the first and the second
moment of the spatial distribution for different values of $Da$.
(a) autocatalytic, (b) bi-stable.}
\end{figure}

\begin{figure}
\caption{The time dependence of the total concentration 
for different values of $Da$, (a) autocatalytic reaction 
and (b) bistable model. The dashed line indicates
the rate of growth of the length of a filament due to advection. }
\end{figure}

\begin{figure}
\caption{Same es Fig5. with rescaled time.
The dashed line shows the time-dependence for the spatially
homogeneous system.}
\end{figure}

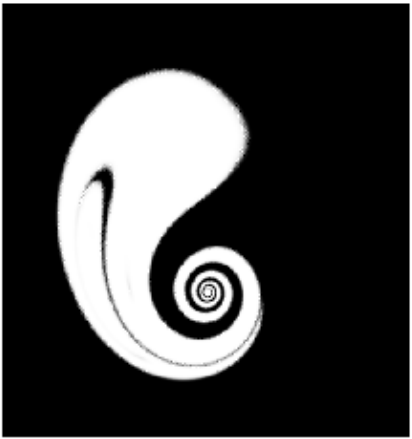
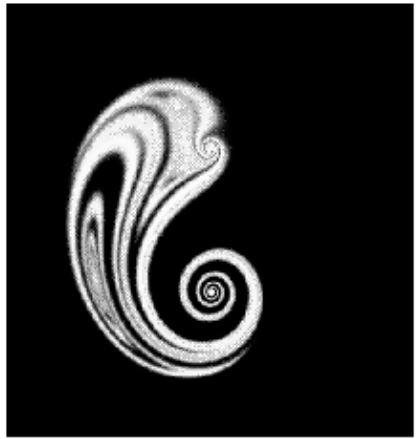
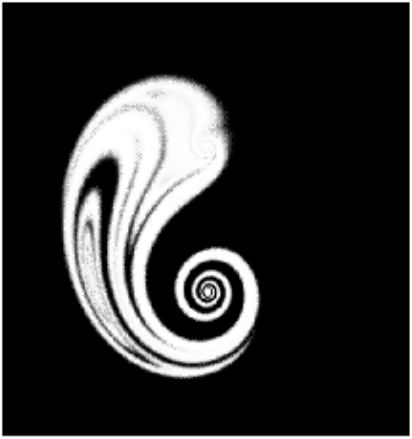
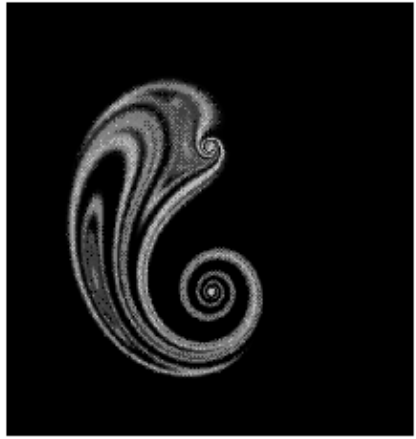
\begin{figure}
\caption{Snapshots of the spatial distribution for the autocatalytic
reaction in the open flow system for a supercritical Damk\"ohler number,
$Da=14.0 > Da_c$ at $t=0.0, 1.0, 2.0, 3.0, 4.0, 5.0$. The distribution
at $t=5.0$ has already reached the time-periodic stationary state.
The amplitude of the perturbation is $C_0=0.5$. For subcritical Damk\"ohler
numbers the perturbation dies out. (The bi-stable model shows similar
behavior but for different values of $Da$.)}
\end{figure}

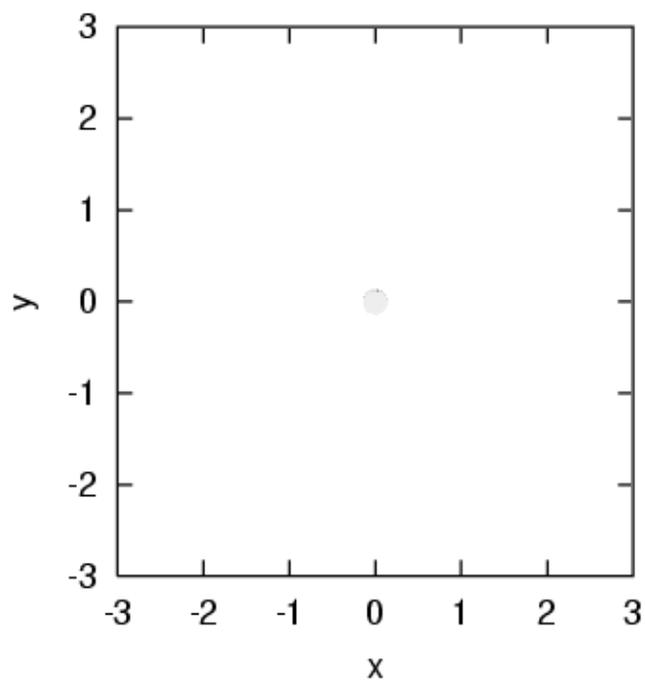
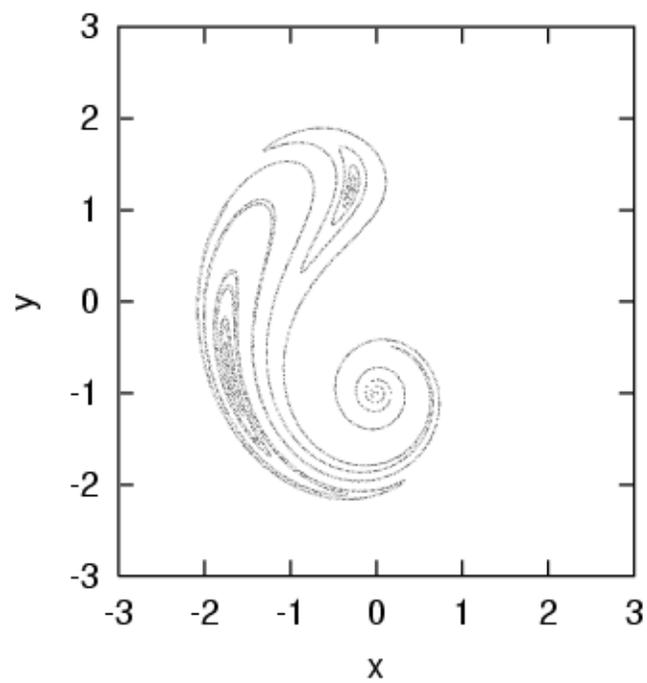
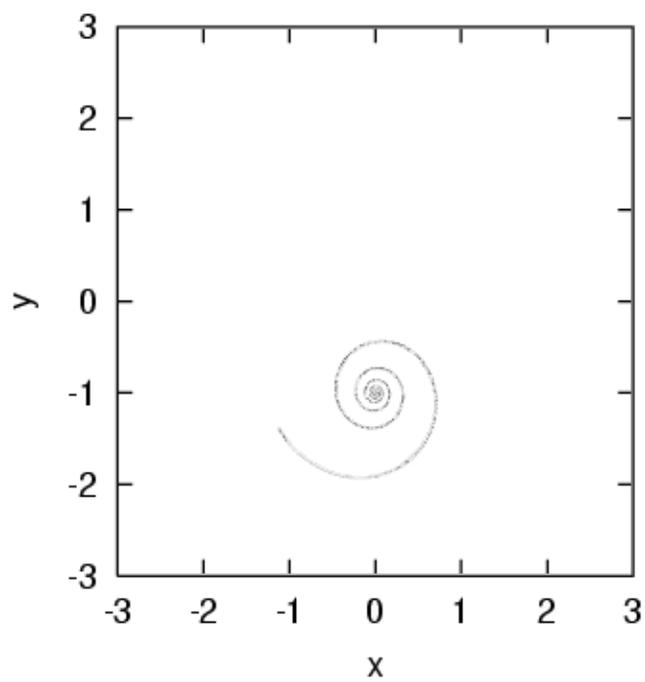
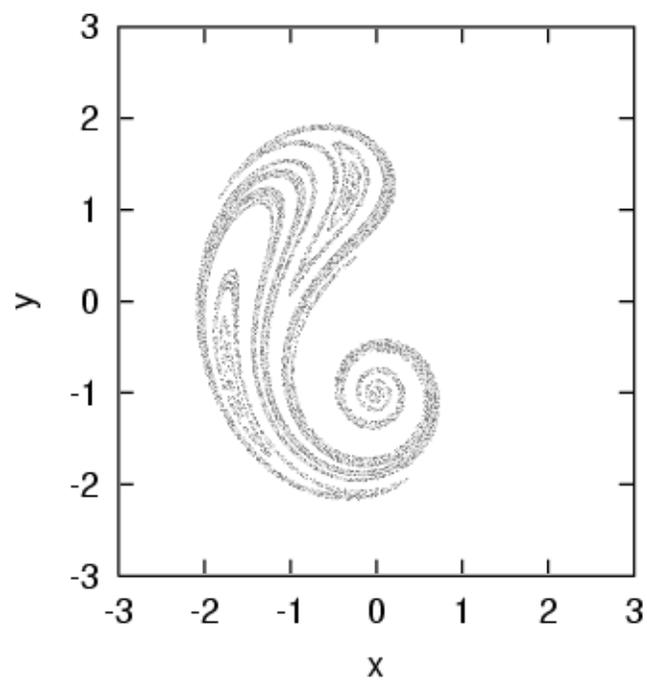
\begin{figure}
\caption{The evolution of an ensemble of particles (e.g. representing a droplet of
dye) in the open flow system ($t=0.0, 2.0, 4.0, 6.0$).}
\end{figure}

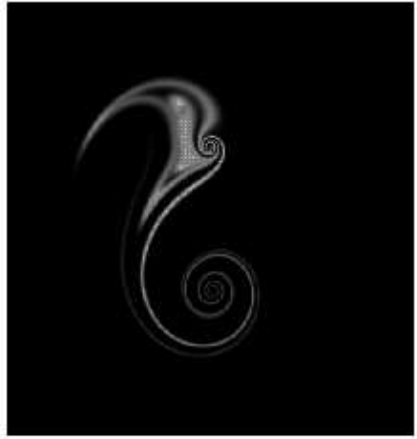
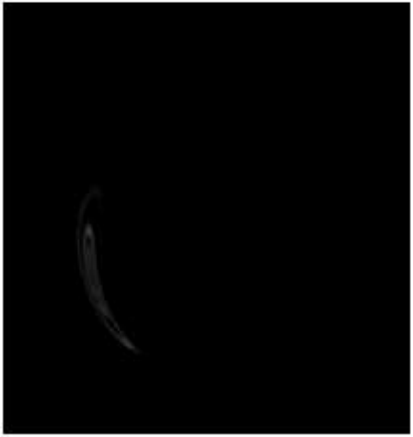
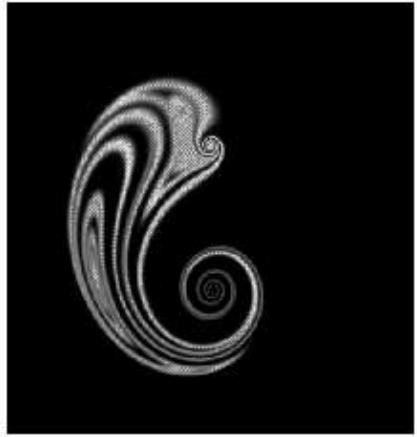
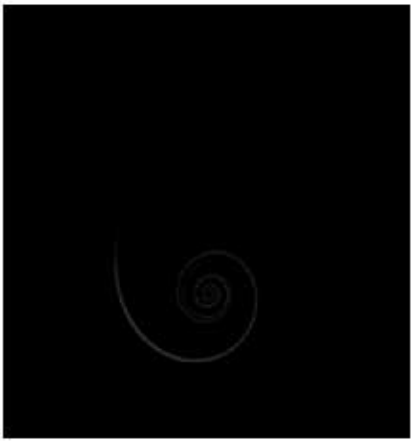
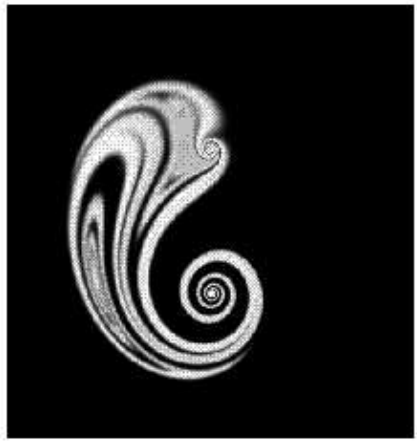
\begin{figure}
\caption{Spatial distribution in the open flow
for the autocatalytic model in the time-periodic 
stationary state for $Da=70.0, 24.0, 11.8, 4.0$.}
\end{figure}

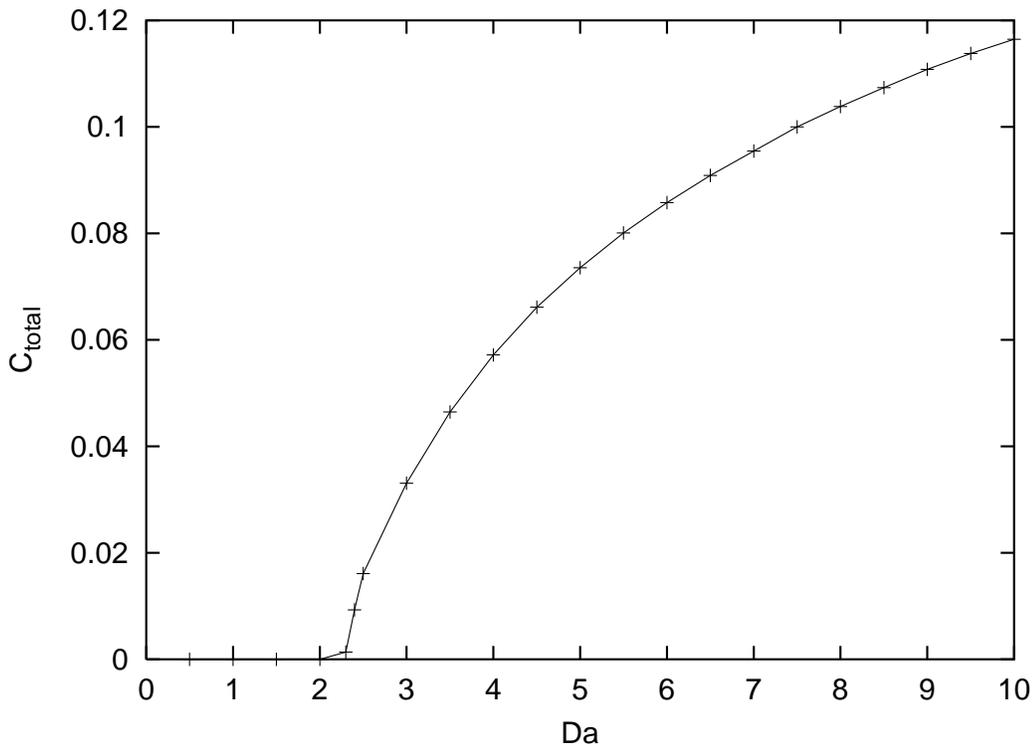
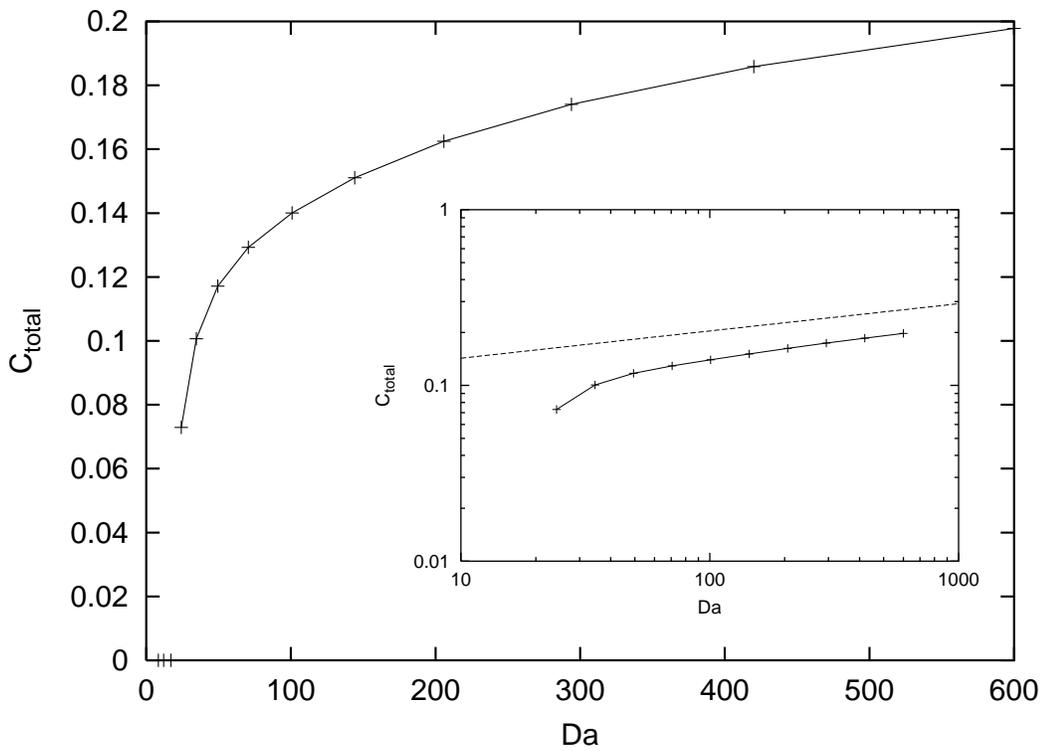
\begin{figure}
\caption{The total concentration in the steady state for the
open flow system as a function of the Damk\"ohler number
for the autocatalytic reaction (a) and the bi-stable model (b).
The dashed line in the inset indicates the predicted asymptotic behavior
$C_{total} \sim Da^{(2-D)/2}$ where $D$ is the fractal dimension of the
unstable foliation of the non-escaping set, $D=1.69$}
\end{figure}

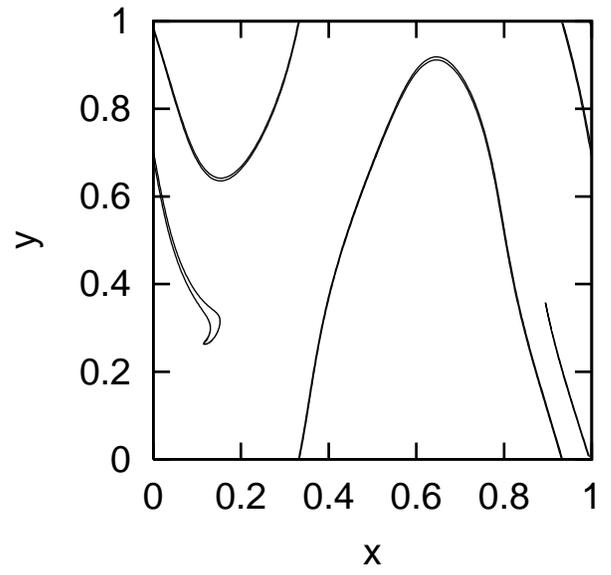
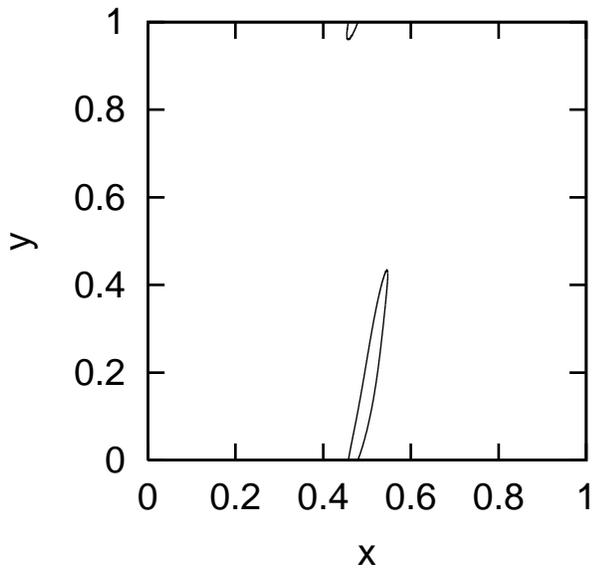
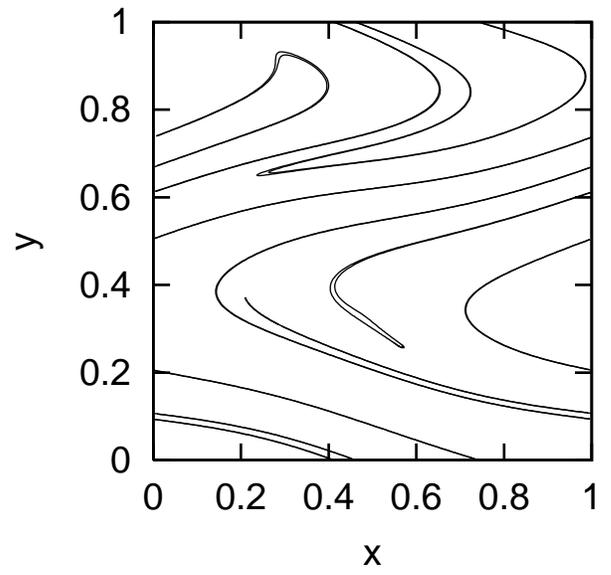
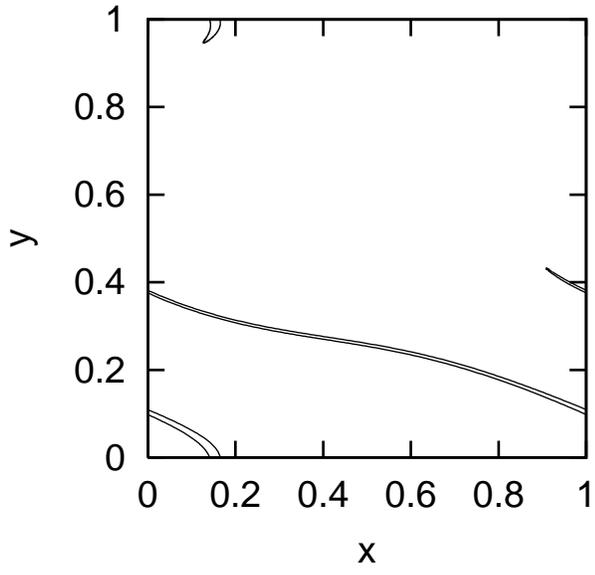
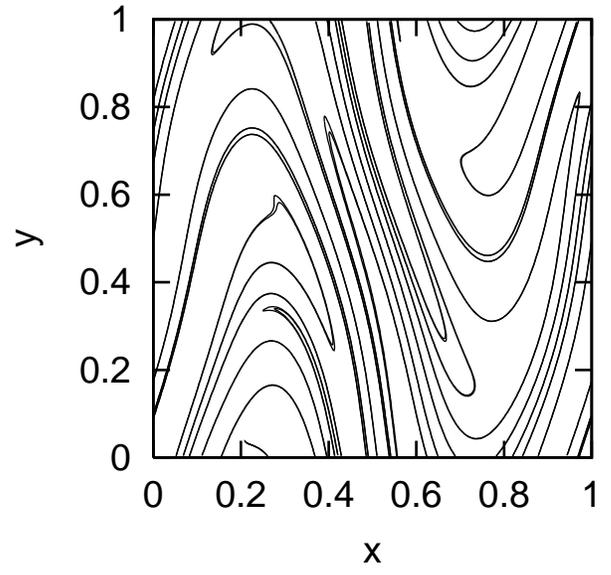
\begin{figure}
\caption{Temporal evolution of a material line advected by the
closed flow. The radius of the initial circular contour is $r=0.06$ and
the figures correspond to $t=0.0, 0.5, 1.0, 1.5, 2.0, 2.5$ as
in Fig.2, for comparison.}
\end{figure}

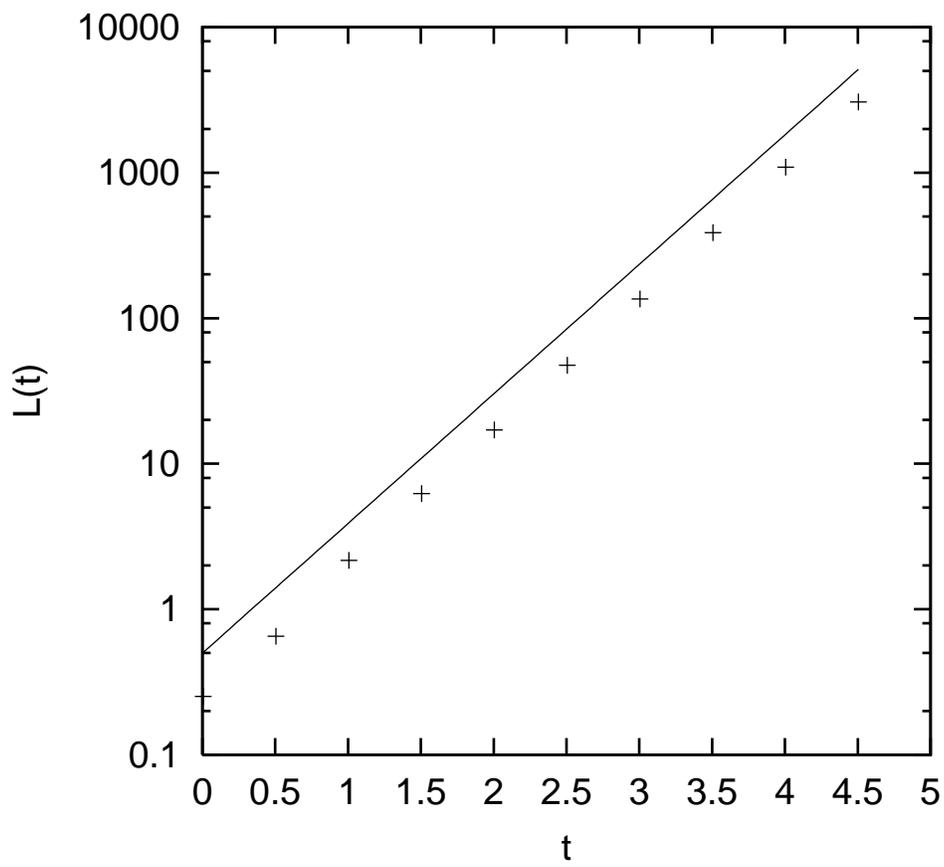
\begin{figure}
\caption{The growth of the length of the contour shown in Fig.11 as a
function of time. The continuous line represents an exponential
growth $\sim \exp{(2.05 t)}$.}
\end{figure}

\begin{figure}
\caption{Steady solutions of the one-dimensional reaction-advection-diffusion
problem, for different values of the Damk\"ohler number 
(a. $\tilde {Da}=80,40,20,10,5.0,2.5,1.25$; 
c. $\tilde {Da}=800,400,200,100,50,25,12.5,11.0$)
and the dependence
of the total concentration in the steady state as a function of $\tilde {Da}$, 
(a,b) autocatalytic, (c,d) bi-stable.
The dashed line in the inset indicates the predicted asymptotic behavior 
$C_{total} \sim \sqrt{\tilde {Da}}$.} 
\end{figure}

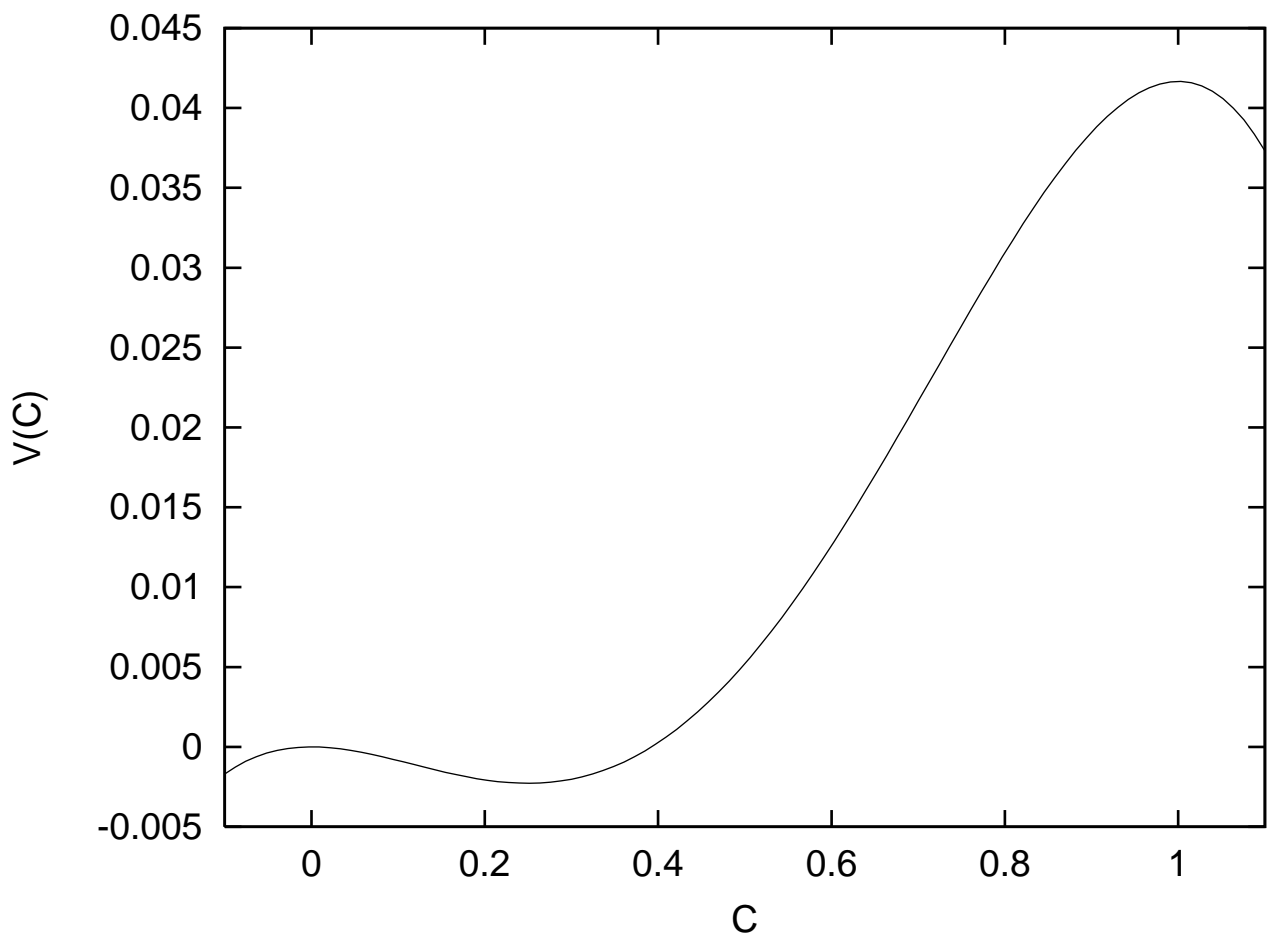
\begin{figure}
\caption{The two-humped potential $V(C)$ for $\tilde {Da}=1$ and $\alpha = 0.25$.}
\end{figure}

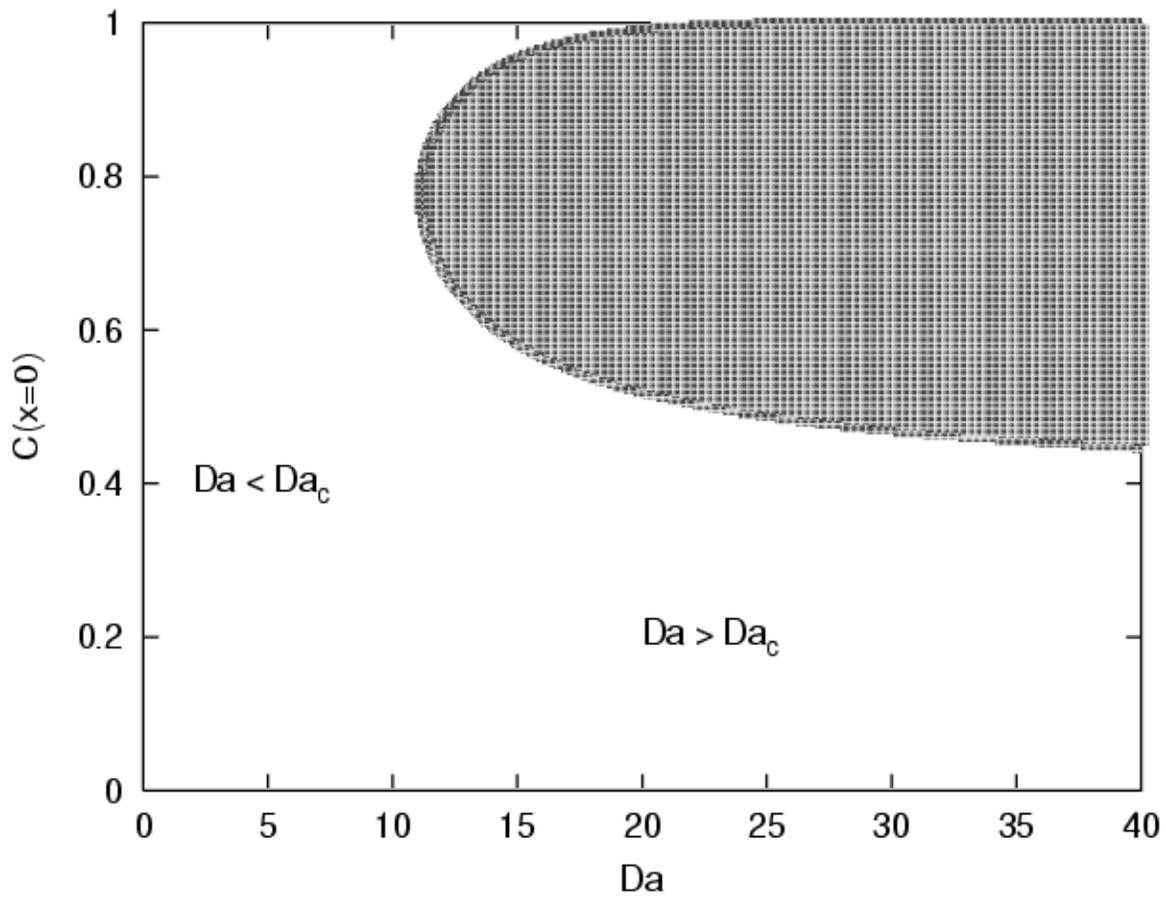
\begin{figure}
\caption{The shaded area shows initial conditions resulting in
the escape of the particle from the potential well. The boundary
of the shaded area corresponds to the steady filaments solutions.
Note, that the filament solution disappears for sub-critical Damk\"ohler
numbers ($\tilde {Da} = 11.0$)}
\end{figure}

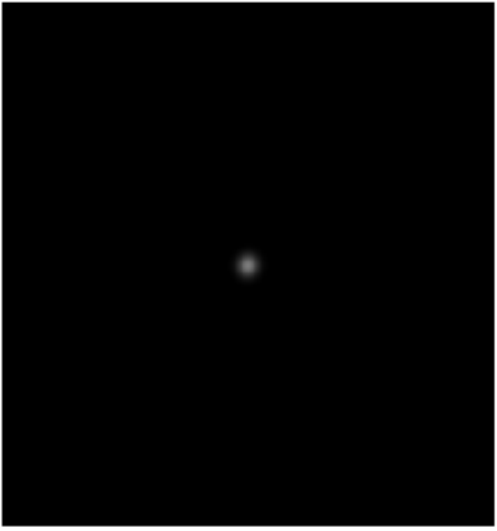
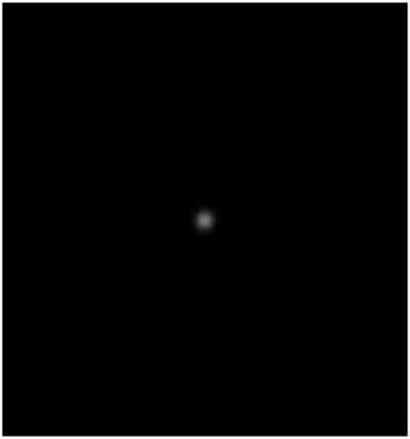
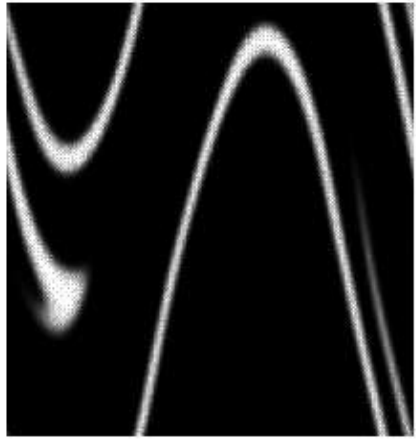
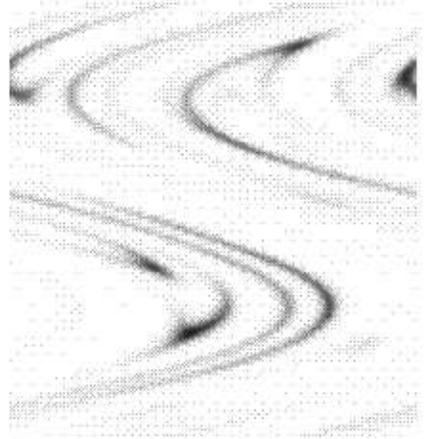
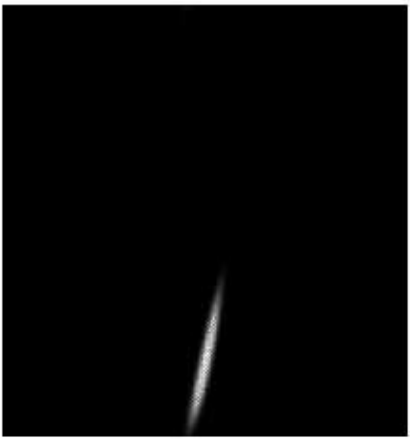
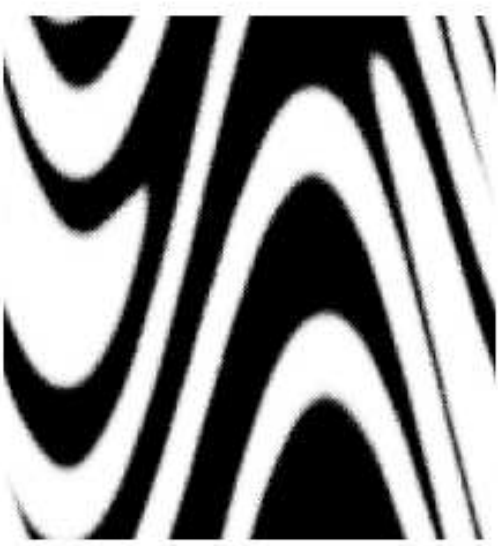
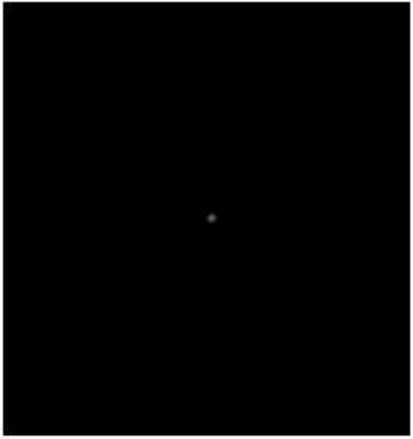
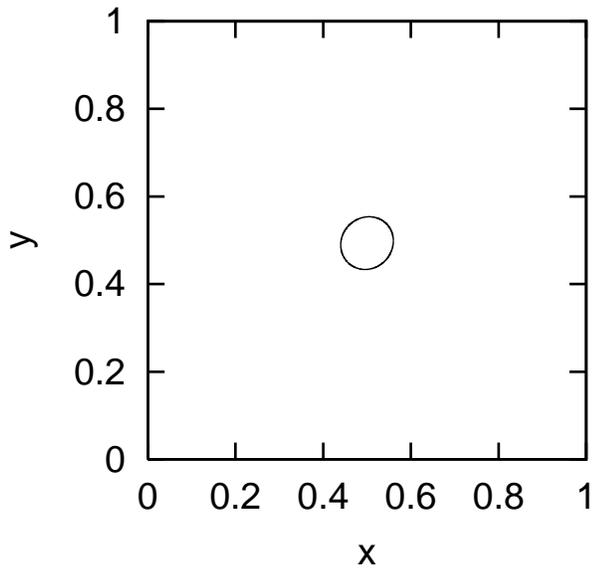
\end{document}